\documentclass[aps,groupedaddress,superscriptaddress,amsmath,amssymb,twocolumn,prb]{revtex4-2}
\usepackage{bm}
\usepackage[dvips]{graphicx}
\usepackage{latexsym}
\usepackage{amsthm}
\usepackage{amsfonts}

\usepackage[colorlinks=true,citecolor=blue,linkcolor=red,urlcolor=blue, linktocpage=true,breaklinks=true,pagebackref=false]{hyperref}
\usepackage{textcomp} 		%\textmu etc
\usepackage{hyperref} 		%clickable links
\usepackage{ifthen} 			% provides \ifthenelse test
\usepackage{xifthen} 		% provides \isempty test
\usepackage{color} 			%colored
\usepackage{soul} 			%highlights with }
\usepackage{lineno}
\usepackage[normalem]{ulem}

\usepackage{xcolor} 
\usepackage{float}

\usepackage{braket}

\def\st{_\mathrm{st}}

\newcommand{\uproman}[1]{\uppercase\expandafter{\romannumeral#1}}

\newcommand{\addMD}[1]{\textcolor{blue}{#1}}

\begin{document}
	
%\pagenumbering{arabic}

\title{Resonant nonlinear response of a nanomechanical system with broken symmetry}
\author{J.\,S. Ochs}
\thanks{Formerly Huber}
\affiliation{Department of Physics, University of Konstanz, 78457 Konstanz, Germany}
\author{G. Rastelli}
\affiliation{INO-CNR BEC Center and Dipartimento di Fisica, Universit{\`a} di Trento, 38123 Povo, Italy}
\author{M. Seitner}
\affiliation{Department of Physics, University of Konstanz, 78457 Konstanz, Germany}
\author{M.\,I. Dykman}
\affiliation{Michigan State University, East Lansing, MI 48824, USA}
\email{dykmanm@msu.edu}
\author{E.\,M. Weig}
\affiliation{Department of Physics, University of Konstanz, 78457 Konstanz, Germany}
\affiliation{Department of Electrical and Computer Engineering, Technical University of Munich, 80333 Munich, Germany}
\email{eva.weig@tum.de}

	\begin{abstract} %(max. 500 words)
		
We study the response of a weakly damped vibrational mode of a nanostring resonator to a moderately strong resonant driving force. Because of the geometry of the experiment, the studied flexural vibrations lack inversion symmetry. As we show, this leads to a nontrivial dependence of the vibration amplitude on the force parameters. For a comparatively weak force, the response has the familiar Duffing form, but for a somewhat stronger force, it becomes significantly different. Concurrently there emerge vibrations  at twice the drive frequency, a signature of the broken symmetry. 
Their amplitude and phase allow us to establish the cubic nonlinearity of the potential of the mode as the mechanism responsible for both observations. The developed theory goes beyond the standard rotating-wave approximation. It quantitatively describes the experiment and allows us to determine the nonlinearity parameters.

	\end{abstract}
	
	\date{\today}
	
	\maketitle
	%\linenumbers

%%%%%%%%%%%%%%%%%%%%%%%%%%%%%%%%%%%%%%%%
%%%%%%%%%%%%%%%%%%%%%%%%%%%%%%%%%%%%%%%%
%%%%%%%%%%%%%%%%%%%%%%%%%%%%%%%%%%%%%%%%
%
%
%	INTRO
%
%	
\section{Introduction} 
Nanomechanical vibrational systems provide a natural platform for studying a broad range of classical and quantum phenomena in a well characterized setting, cf. Refs. \cite{Cattiaux2021,Siskins2021,Tepsic2021,Keskekler2021,Ari2020,MacCabe2020} for recent examples. An important advantageous feature of such systems is a small decay rate, with the ratio $Q$ of the vibration eigenfrequency $\omega_0$ to the decay rate $2\Gamma$ reaching  $5\times 10^{10}$ \cite{MacCabe2020} for localized acoustic modes and $8\times 10^8$ for flexural modes \cite{Ghadimi2018}. This makes nanomechanical modes highly sensitive to a resonant force, which underlies many of their applications. A consequence of this sensitivity is that a comparatively weak resonant force can drive the vibrations into a regime where their nonlinearity comes into play. This enables using resonantly driven nanomechanical vibrations for studying various nonlinear phenomena far from thermal equilibrium, cf. Refs.~ \cite{Aldridge2005,Chan2008a,Segev2008,Defoort2015,Dolleman2019} and papers cited therein. 

In many cases, the nonlinear response of nanomechanical vibrations to a comparatively weak resonant drive is well described by the Duffing model \cite{Kozinsky2007}. In this model the nonlinearity comes from the term in the 
potential energy, which is quartic in the mode coordinate $q$. This is the lowest-order anharmonic term  for a mode with the potential that has inversion symmetry, and in this sense, the Duffing model is minimalistic. The major effect of this term for comparatively small vibration amplitudes comes from making the vibration frequency amplitude-dependent \cite{Landau2004a}. For a small decay rate, the frequency change due to this dependence can exceed the frequency uncertainty due to the decay, making the nonlinearity significant. At the same time, the vibrations remain close to sinusoidal as long as the drive is comparatively weak. The response, in this case,  is often analyzed using the Bogoliubov-Krylov averaging method \cite{Bogoliubov1961}, which in this context is equivalent to the rotating wave approximation (RWA) of quantum optics \cite{Mandel1995}.

Flexural modes, which are most frequently studied in nanomechanics, do not necessarily have inversion symmetry. Such symmetry implies that the nanoresonator lies in a plane and the vibrations occur transverse to this plane. Typically, nanoresonators like nanobeams, nanomembranes, or carbon nanotubes, are bent because of an applied gate voltage \cite{Eichler2013} or the asymmetry of the clamping \cite{Schmid2016} or, as in the system  studied here, because the asymmetry imposed by the dielectric transduction electrodes~\cite{Unterreithmeier2009}, see Fig.~\ref{fig:f1}.  For asymmetric modes, along with the term $\propto q^4$ in the mode potential energy, it is necessary to take into account the lower-order term $\propto q^3$. However, in the standard analysis based on the RWA, the effect of this  term on the vibrations at the drive frequency comes to renormalizing the Duffing parameter \cite{Landau2004a}, such that the standard Duffing model remains applicable with an effective Duffing parameter. 

Here we demonstrate that, for a  nanoresonator with a broken symmetry, the resonant response can significantly deviate from the standard Duffing response already for a moderately strong driving. In the studied system this happens for the vibration amplitudes where the mode frequency differs from its zero-amplitude value $\omega_0$ by $\lesssim 10^{-4}\omega_0$. 

The physics of the effect can be understood from the following argument. In our system the Duffing nonlinearity is hardening,  the frequency increases with the vibration amplitude. On the other hand, the RWA-change of the effective Duffing parameter due to the broken symmetry is negative \cite{Landau2004a}. This means that the symmetry breaking  results in a decrease of the Duffing parameter compared to its value in the symmetric system. For a weak driving it is the decreased effective value that determines the resonant response. However, it is clear that, for sufficiently large amplitudes, the quartic term in the  potential becomes ``stronger'' than the cubic term. Therefore for such amplitudes, the frequency dependence on the amplitude should be different from the small-amplitude range. Remarkably, this happens where the amplitude is still small compared to the scale where the change of the vibration frequency becomes comparable to $\omega_0$. 

Another effect of a broken symmetry is the onset of vibrations at the  even multiples of the frequency of the resonant drive (whereas odd multiples arise from the regular Duffing model), and in particular at twice the drive frequency \footnote{The \addMD{onset of vibrations} \sout{peak} at twice the drive frequency has a counterpart at zero frequency, which was studied for nanoresonators in Ref.~\cite{Eichler2013}.}.  The occurrence of vibrations at twice the drive frequency is often referred to as the second harmonic generation. Such vibrations, which are also referred to as temporal harmonics or overtones, were seen earlier in microscale vibrational systems, cf. Ref.~\cite{Asadi2021} and references therein. Here we measure the amplitude and phase of these vibrations directly, and by comparing them to the amplitude and phase of the main tone establish that they are indeed due to the nonlinearity of the mode potential. This allows quantifying this potential experimentally. 

To describe the observations, the theoretical analysis should go beyond the standard RWA. A simplifying factor is the small decay rate of the nanoresonator studied in the experiment. This suggests extending the methods of the Hamiltonian  nonlinear dynamics to the problem at hand \cite{Arnold1989}. Such an extension should allow for both the nonlinearity and the weak damping. A theory should also address the observation that, even though the signal at twice the drive frequency has an appreciable amplitude in the studied range, higher-order overtones remain very small.

Below in Sec.~II we discuss the setup of the experiment and in Sec.~III summarize the experimental observations. In Sec.~IV we outline the theory. Section~V provides a discussion of the results and a comparison between the theory and the experiment. Section~VI contains concluding remarks. The Appendices describe auxiliary experimental observations, including the observed dependence of the nonlinearity parameters on the applied DC control voltage, a discussion of potential other mechanisms leading to a vibration  at twice the drive frequency, and further details of the theory.

%%%%%%%%%%%%%%%%%%%%%%%%%%%%%%%%%%%%%%%%%%%%%%%%%%%%%%%%%%%%%%%%%%
%
\section{Setup and Characterization}
\begin{figure}[t!]
	\centering
	\includegraphics[width=0.7\linewidth]{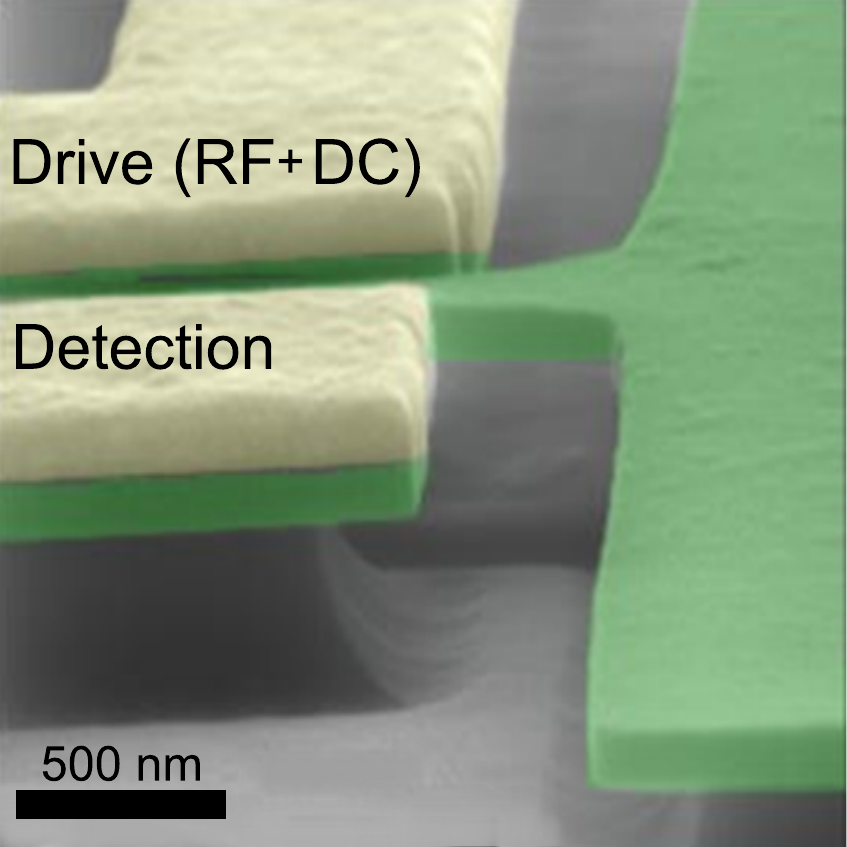}
	\caption{ Scanning electron micrograph of the doubly clamped silicon nitride sting resonator (green) and two adjacent gold electrodes (yellow) for dielectric drive and detection. Depicted is one of the clamping pads (right), the onset of the string and the two control electrodes.}
	\label{fig:f1}	
\end{figure}

We investigate a nanomechanical doubly clamped string resonator fabricated from pre-stressed silicon nitride on a fused silica substrate, similar to the one depicted in Fig.~\ref{fig:f1}. 

%		FIGURE 1

The string is 270\,nm wide, 100\, nm thick and 55\,$\mu$m long.  It is flanked by two adjacent gold electrodes, enabling the dielectric transduction combined with a microwave cavity-enhanced heterodyne detection scheme, discussed in more detail in \cite{Unterreithmeier_2009,Faust_2012,Rieger_2012}. 
The two gold electrodes are placed asymmetrically with respect to the string, leading to an inhomogeneous electrical field when a DC voltage is applied between the electrodes. As a consequence, the dielectric string resonator gets polarized and experiences a gradient force which displaces it from its original equilibrium position as it is getting pulled towards the electrodes where the field is strongest. The corresponding change in the field gradient alters the eigenfrequency, enabling frequency tuning with the applied DC voltage \cite{Rieger_2012}. Concurrently, it also leads to the breaking of the symmetry of the restoring potential (see discussion of Sec.~\ref{sec:theory}). 

For the measurements shown in the following, the DC voltage is fixed to $5$\,V, such that the fundamental flexural out-of-plane mode, which will be considered in the following, is well-separated from the corresponding in-plane mode. 
%We call this mode the M mode. 
The experiment is performed at room temperature of $293$\,K and under vacuum at a pressure below $10^{-4}$\,mbar.

The response of the fundamental out-of-plane mode (referred to as mode M in the following) to a comparatively weak resonant field is described by the Duffing model. In this model the equation of motion for the mode coordinate $q(t)$ reads 
\begin{align}
	\label{eq:Duffing}
	\ddot{q} +2 \Gamma \dot{q} + \omega_0^2 q + \gamma_\mathrm{eff} q^3 = F_{d} \cos(\omega_{d} t)  \, .
\end{align}
Here, $\omega_0=2\pi f_{0}$ is the angular mode eigenfrequency, $\Gamma$ is the damping rate, and $\gamma_\mathrm{eff}$ is the effective Duffing nonlinearity parameter. The parameter $\gamma_\mathrm{eff}$ describes the resonant response in the range of the driving amplitude $F_d$ where the RWA applies, i.e., this is the Duffing parameter renormalized by the cubic nonlinearity of the potential. The driving amplitude $F_d$ in Eq.~(\ref{eq:Duffing})  is scaled by the mass. The driving angular frequency $\omega_d=2\pi f_{d}$ is assumed to be close to $\omega_0$, with $|\omega_d - \omega_0|\ll \omega_0$.

In the experiment, both the drive tone and the measured signal are voltage signals. Therefore we calibrate the system in units of volts, as discussed in detail in the Supplemental Material of Ref.~\onlinecite{Huber2020}, and apply the driving amplitude as an RF voltage $V_d$.
For $V_d < 1$\,mV the mode dynamics is linear. The spectrum of the linear response in this range,  including the Lorentzian fit, is shown in Fig.~\ref{fig:af1} in the Appendix. We find $f_0=6.528$\,MHz and $2\Gamma/(2\pi)$=20\,Hz, which gives the quality factor $Q\approx 325 000$. 

%%%%%%%%%%%%%%%%%%%%%%%%%%%%%%%%%%%%%%%%%%%%%%%%%%%%%%%%%%%%%%%%%%

\section{Experimental Observations}

\begin{figure}[t!]
	\centering
	\includegraphics[width=0.8\linewidth]{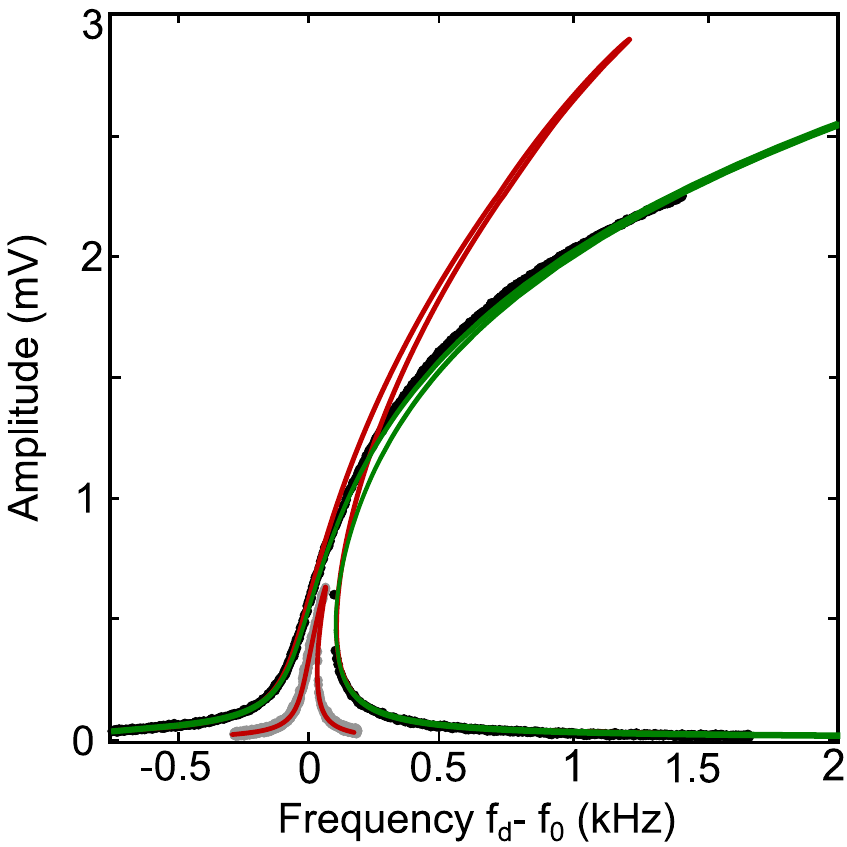}
	\caption{Duffing response and non-Duffing response. Measured nonlinear response curves at drive voltages of $V_d= 9$\,mV (gray) and $40$\,mV (black). Both traces display unprocessed raw data; the noise level remains below the size of the dots. A fit to the Duffing model at $V_d= 9$\,mV as well as a theory curve using the obtained $\gamma_\mathrm{eff}^{(\rm V)}/(2\pi)^2 = 2.48 \cdot 10^{15}$ V$^{-2}$ s$^{-2}$ for $V_d= 40$\,mV are included (red lines). The deviation of the data from the Duffing model is clearly visible for the $V_d=40$\,mV data. A better agreement is achieved by taking the influence of the cubic nonlinearity of the potential $\beta$ into account (green line). The curve is calculated for $\beta^{(\rm V)}/(2\pi)^2=2.08\cdot 10^{15}$ V$^{-1}$ s$^{-2}$  
	and is truncated at a detuning of $2$\,kHz.}
	\label{fig:f2}
\end{figure}

When the drive voltage is increased but remains close to the linear regime, the measured response is well described by the solution of Eq.~(\ref{eq:Duffing}). The corresponding bidirectional scan at a drive voltage of $9$\,mV is shown in Fig.\,\ref{fig:f2}  as gray dots. A fit  (red line) yields the effective nonlinear Duffing parameter. Since the signals are measured in volts, in what follows we use the superscript $(\mathrm{V})$ to indicate that the appropriate nonlinearity parameters obtained by fitting the signals are also in volts. Using the measured value of the signal $q(t)$ in volts, we obtain $\gamma_\mathrm{eff}^{(\rm V)}/(2\pi)^2 = 2.48 \cdot 10^{15}$\,V$^{-2}$ s$^{-2}$.

The measured response for a stronger drive voltage, $V_d=40$\,mV, is shown in Fig.~\ref{fig:f2} by the black dots. The red curve in this figure, on the other hand, shows the response for this voltage calculated using the above value of $\gamma_\mathrm{eff}^{(\rm V)}$. Equation~(\ref{eq:Duffing}) does not fit the data, indicating that the Duffing model no longer applies for $V_d=40$~mV.

%		FIGURE 2

\begin{figure}[b!]
	\centering
	\includegraphics[width=0.7\linewidth]{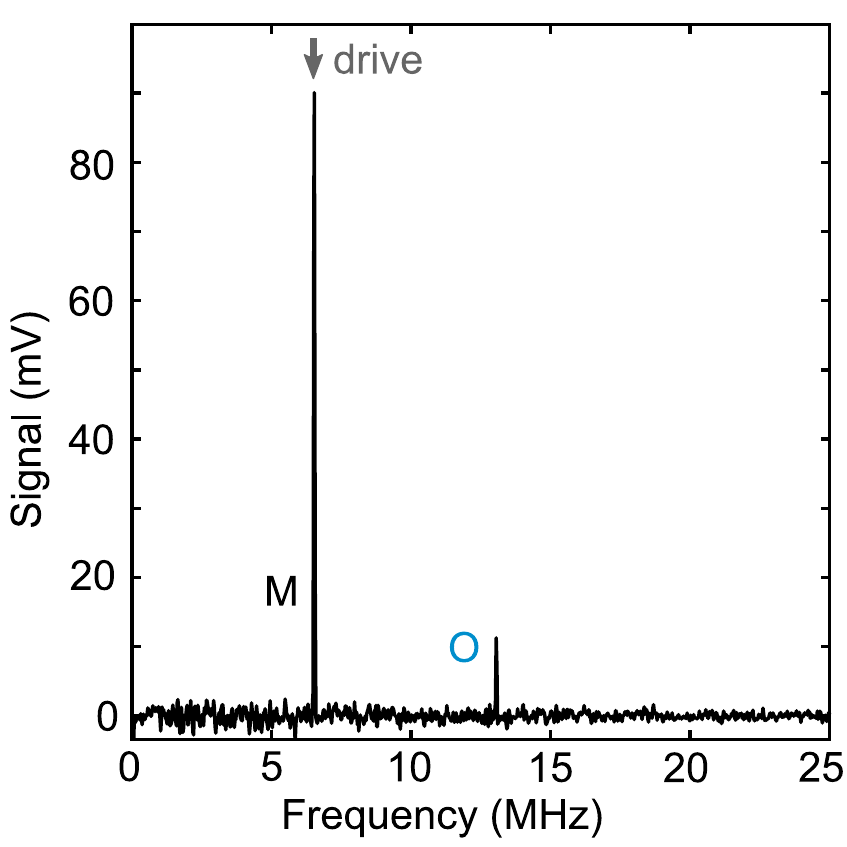}
	\caption{Spectrum of the resonantly driven fundamental mode (M). For a resonant drive at $f_d =f_0=6.528$\,MHz with $V_d=100$\,mV, the  overtone (O) at $13.056$\,MHz is clearly visible. Higher-order overtones are barely discerned in this representation. A constant noise background has been subtracted from the data.}
	\label{fig:f3}
\end{figure}

Along with the deviation from the Duffing response curve we have also observed the onset of a signal at twice the drive frequency. This is depicted for a drive voltage $V_d=100$\,mV applied at the eigenfrequency of the mode, i.e., $f_d = f_0$, in Fig.~\ref{fig:f3}. The spectrum clearly shows the forced vibrations at the eigenfrequency of the driven mode (labelled M), along with a pronounced peak at the  overtone frequency (O) $2f_d =2f_0= 13.056$\,MHz. We note that the second spatial eigenmode of out-of-plane vibrations of the nanostring has the frequency $13.2$\,MHz, almost $200$\,kHz above $2f_d$. A frequency response measurement, indicating the frequency separation of the two features, is shown in Fig.~\ref{fig:af2} in the Appendix. With a frequency separation much larger than the damping rate, both features can clearly be distinguished. This in itself allows us to unambiguously associate the signal overtone at $2f_d$ with the (non-sinusoidal) oscillation of the resonantly driven fundamental mode. 

\begin{figure*}[t!]
	\centering
	\includegraphics[width=0.8\linewidth]{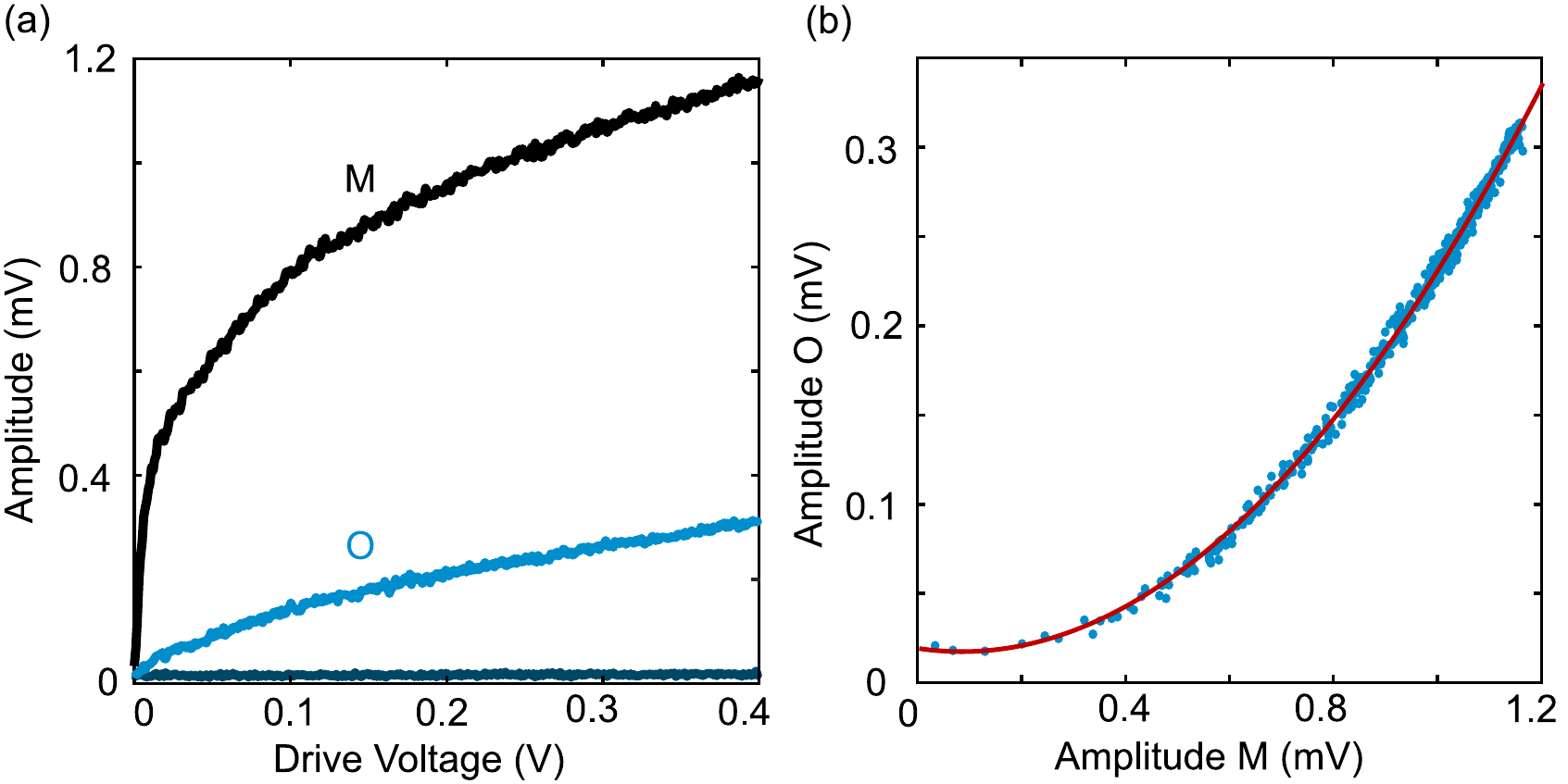}
	\caption{(a) Drive voltage dependence of the resonant amplitude of the fundamental mode (M, black) as well as the amplitudes of the vibrations at twice (O, blue) and three times the drive frequency (dark blue); the amplitudes of the vibrations at the overtones have been rescaled to account for the different frequency-dependent displacement-to-voltage conversion factor.
	The drive frequency $f_d$ is fixed at the resonance frequency of the fundamental mode $f_0$. 
		(b) The amplitude of the vibrations at $2f_d$ [marked by O in panel (a)] depends quadratically on the amplitude of the fundamental mode. Data shown in blue. The red line is a quadratic fit. }
	\label{fig:f4}
\end{figure*}

The dependence on the drive amplitude $V_d$ of the amplitudes of the fundamental mode as well as the signals at $2f_d$ and $3f_d$ are shown in Fig.\,\ref{fig:f4}\,(a). The data refers to the drive frequency $f_d=f_0$. The drive voltage $V_d$ is swept between $0$ and $400$\,mV. 
The three vibration amplitudes are measured simultaneously with a high-frequency lock-in amplifier by using multiple demodulators. 
As already seen in the spectral measurements in Fig.~\ref{fig:f3}, the signal at  $2f_d$ is significantly stronger than the signals at $3f_d$ and $4f_d$ (the signal at $4f_d$ is not shown). 

%		FIGURE 3

%		FIGURE 4

%%%%%%%%%%%%%%%%%%%%%%%%%%%%%%%%%%%%%%%%%%%%%%%%%%%%%%%%%%%%%%%%%%

\section{Theory}
\label{sec:theory}

To account for the broken inversion symmetry, we have to include the cubic nonlinearity in the mode potential $U(q)$ along with the quartic nonlinearity. The equation of motion then reads
\begin{align}
	\label{eq:eom_coordinate}
	\ddot{q} &= - 2 \Gamma \dot{q} -\partial_q U(q)+ F_{d} \cos(\omega_{d} t) \, 
\end{align}
where 
\begin{align}
	\label{eq:potential}
	U(q) = \frac{1}{2}\omega_0^2q^2  + \frac{1}{3}\beta q^3 + \frac{1}{4}\gamma q^4.
\end{align}
Here $\beta$ and $\gamma$ are the parameters of the cubic and quartic nonlinearity, respectively. 
In what follows, we consider a comparatively weak nonlinearity, so that the nonlinear part of the vibration energy remains smaller than the harmonic part $\sim \omega_0^2\langle q^2\rangle$. For concreteness we set $\beta>0$; the sign of $\beta$ can be changed by changing the sign of the coordinate $q$ and incrementing the phase of the drive by $\pi$. We note that in the engineering literature Eq.~(\ref{eq:eom_coordinate}) in the absence of the driving is sometimes called the Helmholtz-Duffing equation. In the context of elastic cables a numerical analysis of Eq.~(\ref{eq:eom_coordinate}) was done  in Ref.~\onlinecite{Benedettini1987}.

For a resonantly driven nonlinear oscillator, a major consequence of the broken symmetry is the occurrence of vibrations at even multiples of the drive frequency beyond the odd multiples expected for a Duffing oscillator. For a weak nonlinearity, forced vibrations are almost sinusoidal,  $q(t) \simeq  q^{(1)}(t) $ with $q^{(1)}(t)  = A\cos(\omega_d t +\varphi)$. To the leading order, the cubic nonlinearity $\beta$  leads to the onset of vibrations at $2{}f_d$. They are described by the expression 
\begin{align}
	\label{eq:second_harmonic}
	q^{(2)}(t) =& \frac{1}{6\omega_0^2}\beta A^2 \cos(2{}\omega_d t + 2 \varphi).
\end{align}
We use that  $|\omega_d-\omega_0|\ll \omega_0$ and $\Gamma \ll \omega_0$. An observation of such vibrations is an unambiguous signature of the broken inversion symmetry of the vibrational mode. There are, however, other contributions to the signal at  $2{}f_d$, which are also related to the broken symmetry, as discussed in Appendix~\ref{sec:second_harmonic}.

Besides leading to the onset of vibrations at the even multiples of the drive frequency, the cubic nonlinearity modifies the dependence of the amplitude of forced vibrations on the drive amplitude compared to the Duffing response. The Duffing model (\ref{eq:Duffing}) has been very successful in describing many observations in  nanomechanical systems, and as mentioned in the Introduction, in the majority of cases the analysis was based on the RWA. In the RWA, one changes from the fast oscillating coordinate $q(t)$ and momentum $p(t)$ to slowly varying in-phase and quadrature components, $q(t)  - i\omega_d^{-1}p(t) = [Q(t) - iP(t)] \, \exp(i\omega_dt)$. In the equations for $Q,P$ one then disregards the terms  that oscillate at the frequency $\omega_d$ and its overtones. Then the major effect of the Duffing nonlinearity is the dependence of the mode frequency on the vibration amplitude $A$ \cite{Landau2004a}  
\begin{align}
	\label{eq:frequency_renormalized}
	\omega\to\omega_\mathrm{eff} \approx \omega_0 + \frac{3\gamma}{8\omega_0} \, A^2 \, ,
\end{align}
with $A$ given by the value of $(Q^2 + P^2)^{1/2}$ in the stable vibrational state. 

The RWA is often applied also to a vibrational system with additional cubic nonlinearity $\beta$. In this approximation the response to the resonant field is mapped onto that of the Duffing model (\ref{eq:Duffing}) with the renormalized nonlinearity parameter $\gamma_\mathrm{eff}$ replacing the bare Duffing parameter $\gamma$ \cite{Landau2004a}
\begin{align}
	\label{eq:renormalized_gamma}
	\gamma\to \gamma_\mathrm{eff} =\gamma-\frac{10\beta^2}{9\omega_0^2}\,.
\end{align}

It is seen from Eq.~(\ref{eq:renormalized_gamma}) that the cubic nonlinearity can strongly affect the amplitude dependence of the mode frequency (\ref{eq:frequency_renormalized}). Indeed, if $\gamma>0$, but $\gamma_\mathrm{eff}<0$, even the sign of $d\omega_\mathrm{eff}/dA^2$ changes. This leads to the so-called zero-dispersion behavior \cite{Dykman1990d} (see Ref.~\cite{Soskin2003} for a comprehensive review). In what follows we consider the case $\gamma$ and $\gamma_\mathrm{eff}>0$, which is relevant for the experiment described in this paper.

The calculation outlined below shows, however, that the strong change of $d\omega_\mathrm{eff}/dA^2$ occurs only in the region of comparatively small amplitudes $A$. Indeed, for large amplitudes the term $\propto q^4$ in $U(q)$ becomes more important than the term $\propto q^3$. 
Simple dimensional arguments show that for amplitudes $ A^2 \gtrsim \omega_0^2\gamma_\mathrm{eff}/\gamma^2$, the RWA approximation (\ref{eq:renormalized_gamma}) becomes inapplicable (see Appendix~\ref{app:theory}). For small $\gamma_\mathrm{eff}/\gamma$ this happens where the nonlinear part of the energy $\sim \gamma A^4$ is still small compared to the harmonic part $\sim \omega_0^2A^2$. 

The analysis of the resonant  response beyond the renormalization~(\ref{eq:renormalized_gamma}) is significantly simplified in the case of weak damping, where the decay rate $\Gamma\ll \omega_0$. Here it is convenient to change from the coordinate and momentum of the mode to its action-angle variables. This is a canonical transformation. For an isolated mode with the Hamiltonian 
\begin{align}
	\label{eq:Hamiltonian}
	H=\frac{1}{2}p^2 + U(q)
\end{align}
the action  $I$ and the angle (phase) $\phi$ are defined as $I=(2\pi)^{-1}\oint p\,dq $ and $\phi = \partial_I  \int p\,dq $ \cite{Landau2004a}. The natural vibration frequency of the mode is 
\[\omega(I)=(\partial I/\partial E)^{-1},\]
where $E$ is the mode energy. The coordinate and momentum are functions of $I$ and $\phi$ and are periodic in $\phi$, 
\[q(I, \phi+2\pi) = q(I,\phi), \quad p(I, \phi+2\pi) = p(I,\phi).\]

In the presence of friction and dissipation, the second-order equation of motion (\ref{eq:eom_coordinate}) becomes a set of two first-order equations for $I$ and $\phi$,
\begin{align}
	\label{eq:eom_I_phi_full}
	&\dot I = R \, \partial_\phi q , \quad   \dot\phi = \omega(I) - R\, \partial_Iq \nonumber\\
	&R=-2\Gamma p + F_d\cos\omega_dt \, .
\end{align}
In the stationary regime forced vibrations occur at the drive frequency $\omega_d$. This means that $\dot\phi \approx \omega_d $, if we neglect  terms oscillating at $\omega_d$. The action $I$ in this regime has a time-independent component and components oscillating at $\omega_d$. However, as seen from Eq.~(\ref{eq:eom_I_phi_full}), keeping oscillating terms in the right-hand sides of the equations for $\dot I$ and $\dot \phi$  lead to small corrections to $I$ and $\phi$ for a comparatively weak drive; in particular, the corrections to $I$ are $\sim \Gamma I/\omega_d$ and $~|F_d \partial_\phi q  \, /\omega_d|$. The smallness of these corrections is the condition of the applicability of the analysis. 

If we disregard the fast-oscillating corrections, the right-hand sides of the equations for $\dot I$ and $\dot\phi$ can be averaged over the period $2\pi/\omega_d$.  In this approximation the equations for the stationary states  read 
\begin{align}
	\label{eq:working}
	d\overline{I}/dt = \overline{R\, \partial_\phi q}=0, \quad d\overline{\phi}/dt = \omega(\overline{I}) -\overline{R\, \partial_Iq}=\omega_d
\end{align}
where the overbar implies period averaging. It is clear from Eq.~(\ref{eq:working}) in particular that only the component of $q(I,\phi)$ that oscillates as $\cos\phi$, i.e., the main tone, contributes to the terms that multiply $F_d\cos\omega_dt$. 

Equation~(\ref{eq:working}) gives two parameters of the stationary vibrational state, the action $\overline{I}=I_\mathrm{st}$ and the time-independent part $\varphi\equiv \varphi_\mathrm{st}$ of the phase, $\overline{\phi(t)} = \omega_d t + \varphi_\mathrm{st}$. The solution of Eq.~(\ref{eq:working}) is provided in Appendix \ref{app:theory}. The calculation is significantly simplified by the fact that the functions $q(I,\phi), p(I,\phi)$ can be expressed in terms of the Jacobi elliptic functions. This property allows one to find the frequency $\omega(I)$ as well as the amplitudes of  vibrations at $\omega(I)$ and its overtones as functions of $I$ in terms of the elliptic integrals. Inversely, for not too strong nonlinearity, it allows one to express the action $I$ in terms of the amplitude $A$ of the main tone, i.e., of the vibrations at frequency $\omega(I)$. 

\begin{figure*}[t!]
		\centering
		\includegraphics[width=0.8\linewidth]{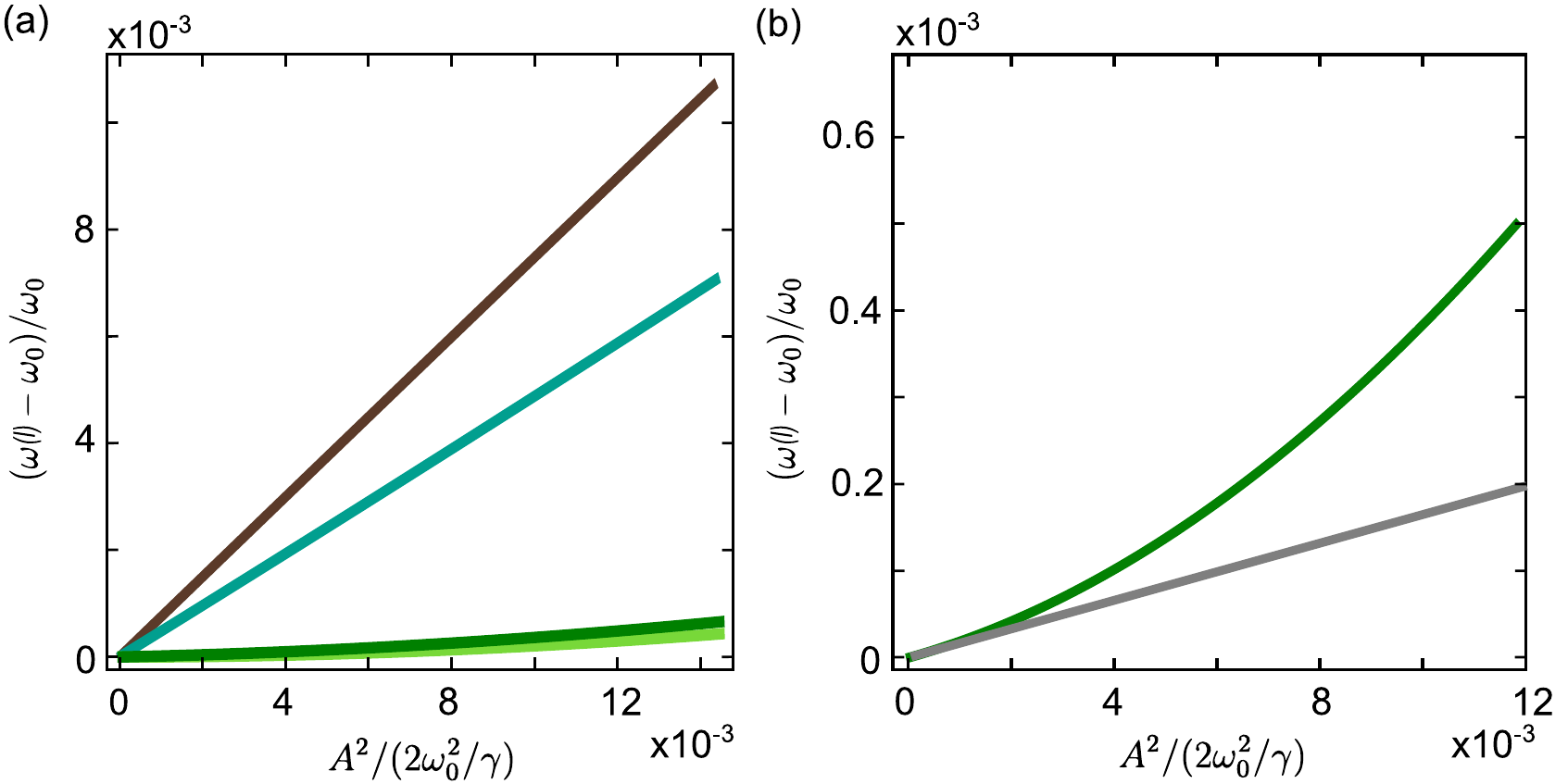}

				\caption  {Relative variation of the frequency $\omega(I)$ vs the square of the amplitude $A$ scaled by $q_0 = \sqrt{2 \omega_0^2 /\gamma}$. 
				(a) The brown, petrol, and light green curves refer, respectively, to $\beta /\omega_0\sqrt{\gamma}= 0, 0.6$, and $\sqrt{0.9}$ (this is the critical value for which $\gamma_\mathrm{eff}=0$). The dark green curve is computed using the value of  $\beta /\omega_0\sqrt{\gamma}$ used in the comparison with the experiment in Fig.~\ref{fig:f2}.
				(b) Closeup showing  the relative variation of $\omega(I)$ using the same $\beta$ as in Fig.~\ref{fig:f2} (dark green) compared with the  Duffing model for the experimental value of $\gamma_\mathrm{eff}$ (gray) for the upper vibrational branch. 
						}
		\label{fig:f5}
	\end{figure*}

In Fig.~\ref{fig:f5}\,(a) we plot the frequency $\omega(I)$ vs the square of the amplitude $A$ scaled by  the typical displacement $(2\omega_0^2/\gamma)^{1/2}$ at which the Duffing nonlinearity becomes pronounced. The plots refer to different values of the scaled cubic nonlinearity  $\beta/\omega_0\sqrt{\gamma}$. For $\beta=0$ (brown line) the frequency is linear in $A^2$ for small amplitudes.
In contrast, for the critical value $\beta_\mathrm{cr}/\omega_0  \sqrt{\gamma} = \sqrt{0.9}$ (light green line), where the effective Duffing parameter $\gamma_\mathrm{eff}=0$, the frequency is parabolic in $A^2$ for small $A^2$. For intermediate $0 <\beta < \beta_\mathrm{cr}$ (petrol line) the frequency displays a significant curvature as function of $A^2$ in the small-$A^2$ range. 
The curve with the experimental value of $\beta$ obtained in Section~\ref{results} (dark green line) is very close to the line for the critical $\beta$, as the effective Duffing parameter is much smaller than the bare Duffing parameter. 
A zoom into the small amplitude regime is shown in Fig.~\ref{fig:f5}\,(b) where 
the experimental range of the solution for the experimental value of $\beta$ (dark green line) is  
compared to the solution of the renormalized Duffing model, Eq.~(\ref{eq:renormalized_gamma}). The two curves clearly disagree.

%To compare the results to the experiment, we convert the vibration amplitude found from Eq.~(\ref{eq:working}) to the output voltage using the procedure described in the Supplemental Material of Ref.~\onlinecite{Huber2020}. From the experimental results for the \sout{second} overtone \addEW{at $2{}f_d$} we found that 
%\addGR{$\beta \sqrt{\gamma}/\omega_0 \lesssim \sqrt{ 0.9}$}, close to but smaller than the critical value $\beta_\mathrm{cr}$.  
%\addJa{Jana: $\beta_\mathrm{cr} \approx 0.94$}

We  note that a simple way to think of the cubic term in the potential of a nanoresonator mode is to relate it to a linear bias. Such bias can come, for example, from a gate voltage that ``pulls'' the nanoresonator. The  potential of the Duffing oscillator with linear bias is
\begin{align}
	\label{eq:bias}
	U_B(q) = -\xi_\mathrm{bias} q + \frac{1}{2}\omega_0^2 q^2  + \frac{1}{4}\gamma q^4,
\end{align}
where $\xi_\mathrm{bias}$ is the bias strength.

The potential $U_B$ has the same form as  the potential~(\ref{eq:potential}) for $|\beta|<(4\gamma\omega_0^2)^{1/2}$. This is seen if one shifts the equilibrium position $q\to q+\delta q$ to compensate the linear term; in the limit of small nonlinearity $\delta q \approx \xi_\mathrm{bias}/\omega_0^2$. After the shift $U_B(q)$ becomes of the same form as $U(q)$ with 
\[\beta = 3\gamma \delta q, \quad \omega_0^2 \to \omega_0^2 + 3\gamma (\delta q)^2.\]  
The condition  $\beta^2 < 4 \gamma \omega_0^2$ corresponds to the experimental situation where the potential $U(q)$ has a single minimum.

%					FIGURE 5

%%%%%%%%%%%%%%%%%%%%%%%%%%%%%%%%%%%%%%%%%%%%%%%%%%%%%%%%%%%%%%%%%%

\section{Results}\label{results}

The response curve shown in Fig.~\ref{fig:f2} is anomalous in the sense that it significantly differs from the conventional Duffing curve. The theoretical model of Sec.~\ref{sec:theory}, which takes into account the cubic nonlinearity of the potential, allows us to describe this curve quantitatively and to find the nonlinearity parameter $\beta^{(\rm V)}$. The analysis is based on Eqs.~(\ref{eq:working}) and uses the experimentally determined eigenfrequency $\omega_0$ and linewidth $2\Gamma$ (see  Appendix \ref{app:theory}).

The effective Duffing parameter $\gamma_\mathrm{eff}^{(\rm V)}$ (in volts) is extracted from the measurements for a moderately weak driving, where the effective Duffing model applies, cf. the data for $V_d=9$~mV in Fig.~\ref{fig:f2}. The cubic nonlinearity parameter $\beta^{(\rm V)}$ is then chosen to match the theoretical result to the experimental curve for a stronger drive. The underlying theory goes beyond the conventional RWA approximation.  The green line in Fig.~\ref{fig:f2} displays the theoretical curve for $\beta^{(\rm V)}/(2\pi)^2=2.08 \cdot 10^{15}$\,V$^{-1}$s$^{-2}$. With this parameter value, we find good agreement between the theory and the experiment.

However, it is seen from Fig.~\ref{fig:f2} that the upper branch of the experimental response curve for $V_d=40$~mV ends at a detuning of approximately $1.5$\,kHz, whereas the theoretical branch extends much further and is truncated at a detuning of $2$\,kHz. We attribute this discrepancy to a comparatively short lifetime of the large-amplitude state. Our data acquisition system does not allow us to observe state with a short lifetime, and therefore we do not observe the stable large-amplitude state beyond a certain frequency. 

We now comment on the lifetime of the large-amplitude state.  As seen from Fig.~\ref{fig:f2}, the theoretical values of the amplitudes of this state and the unstable state  are very close. In the considered very weakly damped system these amplitudes are determined by the quasienergies (Floquet eigenvalues) of the periodically driven mode in the corresponding states, which indicates that the corresponding quasienergies are also close. Thermal noise, which is invariably present in the system, leads to escape from a dynamically stable vibrational state \cite{Dykman1979b}. In the system investigated here such escape was seen earlier \cite{Huber2020}. 
%[\addJa{Do you refer to the critical switching? But this was measured in a much slower measurement! I guess we are far off the critical switching point here?}]

In the weak-damping regime, periodically driven systems escape via diffusion over quasienergy. Generically, the escape rate increases exponentially with the decreasing distance between the quasienergies of the stable and unstable (saddle-type) states \cite{Dykman2005d}. Therefore we expect it to be comparatively large where the amplitude of the stable state is close to that of the unstable state. A full calculation of the escape rate is beyond the scope of the present paper.

The values of $\gamma_\mathrm{eff}^{\rm (V)}$ and $\beta^{\rm (V)}$ allow us to find the ``bare'' Duffing parameter $\gamma^{(\rm V)}$ using Eq.~(\ref{eq:renormalized_gamma}). The value of this parameter  $\gamma^{(\rm V)}/(2\pi)^2 \approx 1.16 \cdot 10^{17} \, \rm V^{-2} \rm s^{-2} $  is two orders of magnitude larger than the effective Duffing coefficient $\gamma_\mathrm{eff}^{(\rm V)}/(2\pi)^2 = 2.48 \cdot 10^{15}$ V$^{-2}$s$^{-2}$ measured for the moderately weak driving regime. 
	
The signal observed in the experiment at twice the drive frequency arises from the broken inversion symmetry of the nanostring under investigation. The symmetry breaking is caused primarily by the asymmetric arrangement of the dielectric transduction electrodes (Fig.~\ref{fig:f1}). However, besides the mechanism described in the previous section and associated with the cubic nonlinearity of the potential $\propto \beta$, other mechanisms originating from the broken inversion symmetry could also contribute. They include the nonlinear excitation of the second spatial harmonic eigenmode of the nanostring and the nonlinear coupling to the driving force resulting in a direct as well as a parametric drive at $2f_d$. These mechanisms are discussed in more detail in Appendix~\ref{sec:second_harmonic}. Notably, some of them lead to a different dependence of the signal at $2f_d$ on the amplitude and phase of the fundamental mode M. This can be used to identify the origin of the signal. In the following, we compare our experimental observations with the theoretical predictions for all these mechanisms and indeed find that the nonlinearity of the potential energy of the nanoresonator characterized by the parameter $\beta$ is the dominant source of the signal at $2f_d$.

Figure\,\ref{fig:f4}\,(b) displays the amplitude of the  overtone signal (in volts) $V_{\rm out,O}$ at $2{}f_d$ as a function of the amplitude $V_{\rm out, M}$ of the resonant response of the mode M. A fit with $V_{\rm out,O}=cV^2_{\rm out, M}$ clearly shows that the overtone amplitude scales quadratically with that of the main tone. In addition, we have determined the phase of the signal measured at $2{}f_d$ with respect to the phase $\varphi$ of the  mode M with the lock-in amplifier. The measurement indeed reveals that the phase of the overtone coincides with $2\varphi$ for a comparatively weak driving  used to derive Eq.~(\ref{eq:second_harmonic}). Both observations are in agreement with Eq.~(\ref{eq:second_harmonic}).
This suggests that the nonlinear driving terms (\ref{eq:2nd_overtone_direct}) and (\ref{eq:nonlinear_suscept}) discussed in Appendix \ref{sec:second_harmonic} make a small contribution to the  signal  at $2{}f_d$ at most, since they exhibit a different dependence on the amplitude and phase of the main tone.

However, this is not sufficient to establish the nonlinearity of the mode potential $U(q)$  as the only source of the signal at $2{}f_d$, as the experimental observations are also compatible with the nonlinear coupling to the second spatial harmonic  of the nanostring, see  Eq.~(\ref{eq:mode_2_vibrations}). In order to estimate the role of this mechanism as the remaining alternative source of the overtone signal, we repeat the experiment for a drive frequency at half the eigenfrequency of the fundamental mode $\omega_0/2$. In this regime, according to Eq.~(\ref{eq:eom_coordinate}), 
the force $F_d\cos(\omega_dt)$ still excites forced  vibrations with a displacement $q^{(0)}(t)\approx [F_d/(\omega_0^2-\omega_d{}^2)]\cos\omega_d t$, even when nonresonant as it is the case for $\omega_d \approx \omega_0/2$. Because of the nonlinearity of the mode, which is characterized by the parameter $\beta$ in Eq.~(\ref{eq:potential}), these vibrations resonantly excite vibrations at $2{}f_d$, which is the effect of resonant second harmonic generation,
	\begin{align}
		\label{eq:half_eigenfrequency}
		%&
		q^{(2)}(t) \approx \frac{\beta F_d{}^2}{2(\omega_0^2-\omega_d{}^2)^2}
		%\nonumber\\
		%&
		\times\mathrm{Re}\frac{\exp(2i\omega_d t)}{\omega_0^2 - 4\omega_d{}^2 +4i\Gamma\omega_d}, \nonumber \\ \quad \omega_d\approx \omega_0/2.
	\end{align}

\begin{figure*}[t!]
	\centering
	\includegraphics[width=0.8\linewidth]{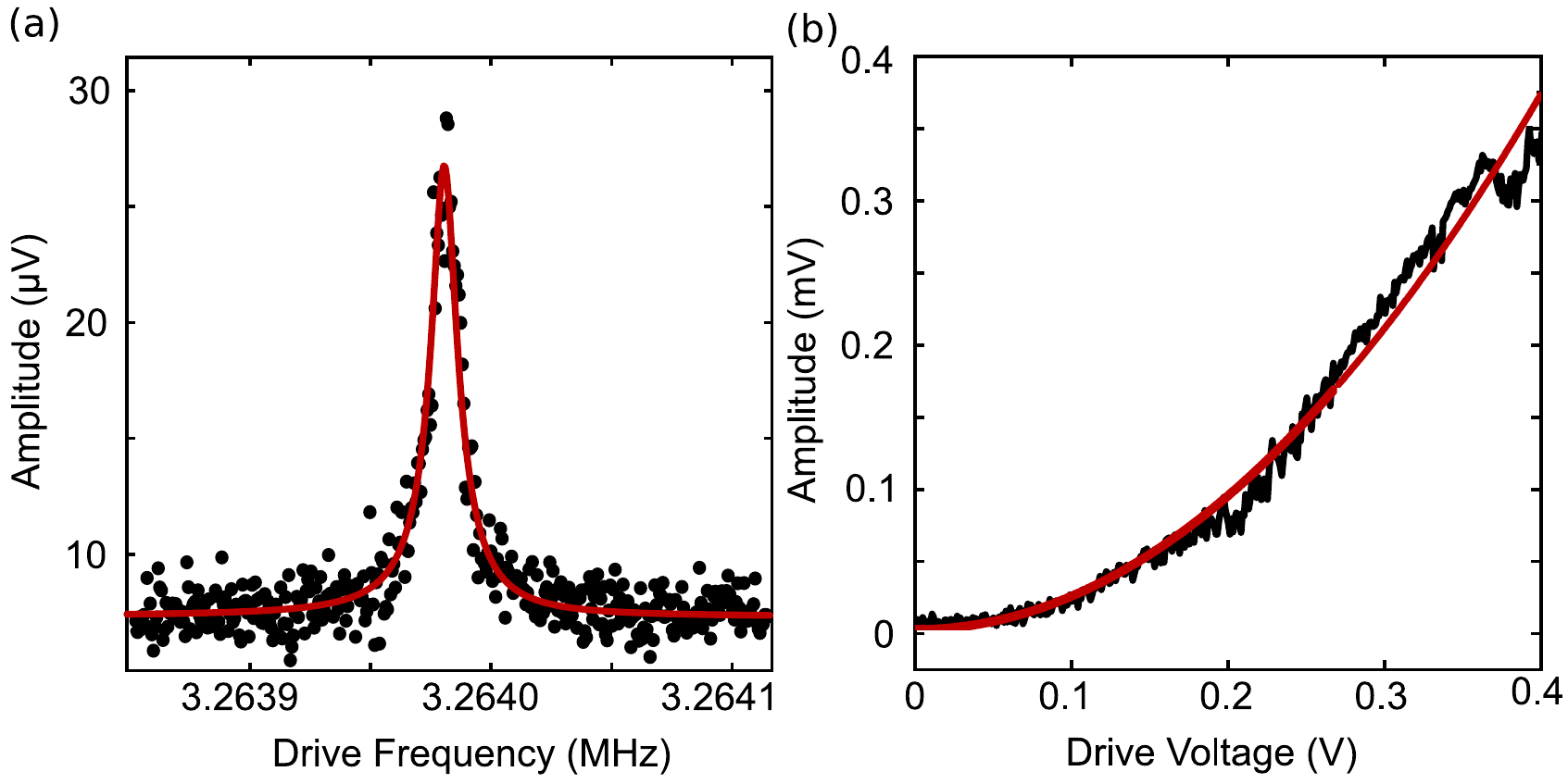}
	\caption  {Resonant excitation of the overtone  for a drive at half the eigenfrequency. (a) Amplitude of the resonantly excited vibrations at $2f_d$ as a function of the drive frequency $f_d$ swept around $f_0/2$ with $V_d=100$\,mV. The red line displays a Lorentzian fit with halfwidth $\Gamma$. (b) Peak amplitude of the resonantly excited vibrations at $f_0$ for $f_d=f_0/2$ as a function of the drive voltage. The red line correponds to a quadratic fit.
	}
	\label{fig:f6}
\end{figure*}

As seen from Fig.~\ref{fig:f6}, we clearly observe the corresponding overtone at $2f_d\approx f_0$, whereas the response at the drive frequency $f_d$ could not be resolved for bandwidth limitations of the experimental setup. 
Figure~\ref{fig:f6}\,(a) displays the overtone amplitude for a sweep of the drive frequency around $f_0/2$ along with a Lorentzian fit. The fit is in full agreement with Eq.~(\ref{eq:half_eigenfrequency}). Figure~\ref{fig:f6}\,(b) shows the scaling of the overtone amplitude for $f_d=f_0/2$ with the drive voltage fitted with a quadratic function.

The response of the second spatial harmonic of the nanostring is nonresonant for $f_d\approx f_0/2$ and may not lead to an appreciable signal. This demonstrates that the cubic nonlinearity of the potential $U(q)$  of the mode M is the major contributor to the response at twice the drive frequency and suggests that it is also a major contributor in the case of the driving at $f_d\approx f_0$. In Appendix \ref{sec:second_harmonic} we provide an argument why the nonlinear coupling to the second spatial harmonic of the nanostring should be weak in addition to being just nonlinear. 
%[\addJa{What is meant by being just nonlinear?}]

%					FIGURE 6

The finding that the observed signal at $2f_d \approx 2f_0$ is predominantly caused by the nonlinearity $\beta$ suggests that the ratio $c=V_{\rm out,O}/V^2_{\rm out, M}$, which can be extracted from the quadratic fit to the data in Fig.~\ref{fig:f4}\,(b), could be employed to quantitatively determine the cubic nonlinearity parameter in units of volts, $\beta^{(\rm V)}$, from Eq. (\ref{eq:second_harmonic}). However, the displacement-to-voltage conversion factor of our read-out apparatus at $2f_0$ does not coincide with the calibrated one at $f_0$ \cite{Gajo_2020}, such that $\beta^{(\rm V)}$ cannot be directly extracted from the data.

%%%%%%%%%%%%%%%%%%%%%%%%%%%%%%%%%%%%%%%%%%%%%%%%%%%%%%%%%%%%%%%%%%
\section{Conclusions}
Conventionally, the nonlinear response of vibrational modes to moderately strong resonant driving is described by the Duffing model. In contrast, in this work we have observed and explained a non-Duffing resonant nonlinear response of  the fundamental mode of an underdamped  nanomechanical resonator. We found that, even though the dependence of the amplitude of forced vibrations on the frequency of the drive is of the familiar Duffing form for moderately weak driving, its shape changes significantly as the driving becomes stronger. This happens in the range where the driving still remains not too strong, so that the forced vibrations are still close to sinusoidal. Our explanation is based on taking into account the broken inversion symmetry of the resonator. The symmetry breaking leads to the onset of a term in the potential energy of the mode, which  is cubic in the mode coordinate. 

We show that, while the cubic term does not change the form of the response to a moderately weak drive, it significantly changes the response for a stronger drive. The analysis required us to go beyond the standard rotating-wave approximation. The obtained results allow describing the spectrum of the nonlinear response in a broad range of amplitudes of the resonant drive, where the shape of the response curve changes significantly. The comparison between the experimental data and the theoretical model allows us to determine the parameters of the quartic and cubic nonlinearity of the potential of the mode.

Along with the change of the response form, the broken inversion symmetry leads to the onset of response at even multiples of the drive frequency. For the drive at frequency close to the mode eigenfrequency, we have observed vibrations at twice the drive frequency, i.e., second harmonic generation, in optics terms. We have discussed several microscopic mechanisms that lead to the onset of such vibrations. The dependence of the vibration amplitude and phase on the amplitude and phase of the vibrations at the drive frequency suggests that the major contribution to the frequency doubling comes from the cubic nonlinearity of the mode potential. Characteristically, higher overtones have a very small amplitude, consistent with the model. 
%[\addJa{The small amplitudes of the overtones is only discussed in the Appendix, right? Should we add a reference to the section?}]

We have also observed resonant second harmonic generation when the driving frequency was close to half the mode eigenfrequency. The spectrum of the vibrations at twice the drive frequency has a characteristic Lorentzian shape, while the vibration amplitude is quadratic in the driving amplitude.

Our results demonstrate that weakly damped vibrations of nanomechanical systems display very rich nonlinear dynamics.  It comes from the interplay and competition of different nonlinearity parameters even in simple cases where the vibrations remain close to sinusoidal.  The significant compensation of the nonlinearity that we have found in a certain range of the vibration amplitudes, which is controlled by how strongly the inversion symmetry is broken, can be used in applications as it extends the practically important regime of almost linear behavior of the mode.  At the same time, the deviation from the traditionally assumed Duffing behavior should be generic for weakly damped nanomechanical modes, as in many cases such modes lack inversion symmetry. %[\addJa{The compensation and control of the nonlinearity is only shown in the Appendix!}]

%analyzed the non-Duffing response function of an nanomechanical resonator with broken inversion symmetry. In a series of experiments we evaluate the amplitude and phase dependence of the associated \sout{second} overtone appearing at twice the drive frequency, which is also a consequence of the broken symmetry, and conclude that the cubic nonlinearity of the potential $\beta$ is responsible for the observed anomalous response function. We further provide a theoretical model going beyond the standard rotating wave approximation based on action-angle variables which is able to describe the measured response yielding excellent agreement. We show that the amplitude of the \addEW{vibrations at twice the drive frequency} \sout{second overtone} can, in principle, be employed to quantitatively determine the cubic nonlinearity $\beta$, provided the detection sensitivity of the experimental setup is known both at $\omega_0$ and $2\omega_0$. Alternatively, we extract the value of $\beta$ from the modified response curve.

%Note that if the Duffing nonlinearity is softening, the cubic nonlinearity leads to an increased effective Duffing parameter for weak drive, and for a stronger drive the Duffing parameter becomes effectively smaller.

\section{Acknowledgements}
We are grateful to H. Yamaguchi for the discussion of the mechanisms of overtone generation. J.\,S.\,O. and E.\,M.\,W. gratefully acknowledge financial support from the Deutsche Forschungsgemeinschaft
(DFG, German Research Foundation) through Project-ID 425217212 - SFB 1432, the European Union’s Horizon 2020 Research and Innovation Programme under Grant
Agreement No 732894 (FET Proactive HOT), and the German Federal Ministry of Education and Research
(contract no. 13N14777) within the European QuantERA cofund project QuaSeRT.
M.\,I.\,D. acknowledges support from the National Science Foundation, Grants No. DMR-1806473 and CMMI 1661618. M.\,I.\,D. is a senior fellow of the Zukunftskolleg of the University of Konstanz; he is grateful for the warm hospitality at the University of Konstanz where this work was started.

%%%%%%%%%%%%%%%%%
%
%
\appendix 
\section{Linear Response}
\begin{figure}[ht!]
	\centering
	\includegraphics[width=0.7\linewidth]{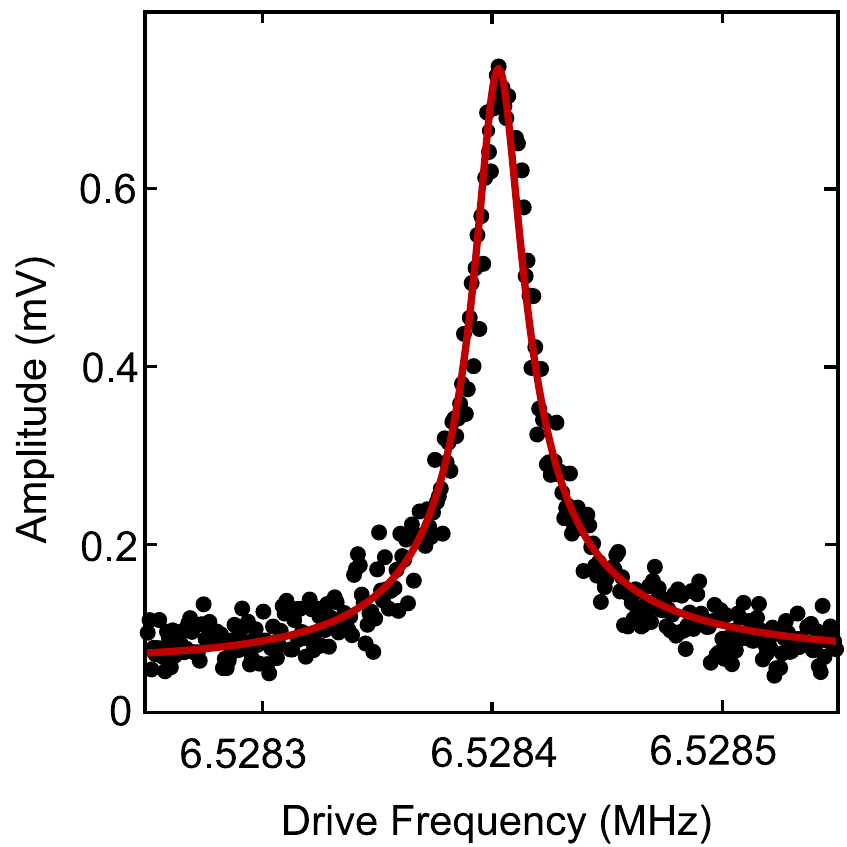}
	\caption{Linear response of the fundamental out-of-plane mode (M) at a drive voltage of $V_{d} =1$\, mV. A Lorentzian fit (red line) yields an eigenfrequency of 6.528\,MHz, a linewidth of $2\Gamma/(2\pi)$=20\,Hz, and a quality factor of $Q\approx 325 000$.}
	\label{fig:af1}
\end{figure}

The linear response of the fundamental out-of-plane mode is found at an eigenfrequency of  $f_0=6.528$\,MHz. It is shown for a drive of $V_{d}=1$\,mV in Fig.~\ref{fig:af1} as black dots along with a Lorentzian fit (red line). From the fit, we extract a linewidth  $2\Gamma/(2\pi)$=20\,Hz, yielding a quality factor of $Q\approx 325 000$. 

%						 FIGURE A1
%%%%%%%%%%%%%%%%%%%
%%%%%%%%%%%%%%%%%%%%

\section{The vibrations at twice the drive frequency}
\label{sec:second_harmonic}

It is tempting to try to extract the value of the nonlinearity parameter $\beta$ from the amplitude of the signal at twice the drive frequency. However, besides the cubic nonlinearity of the mode potential, there are several other mechanisms giving rise to the  generation of the second temporal harmonic. The simplest of them will be discussed in this Section. 

\subsection{Second spatial harmonic of the nanostring}

\begin{figure} [t!]
	\centering
	\includegraphics[width=0.7\linewidth]{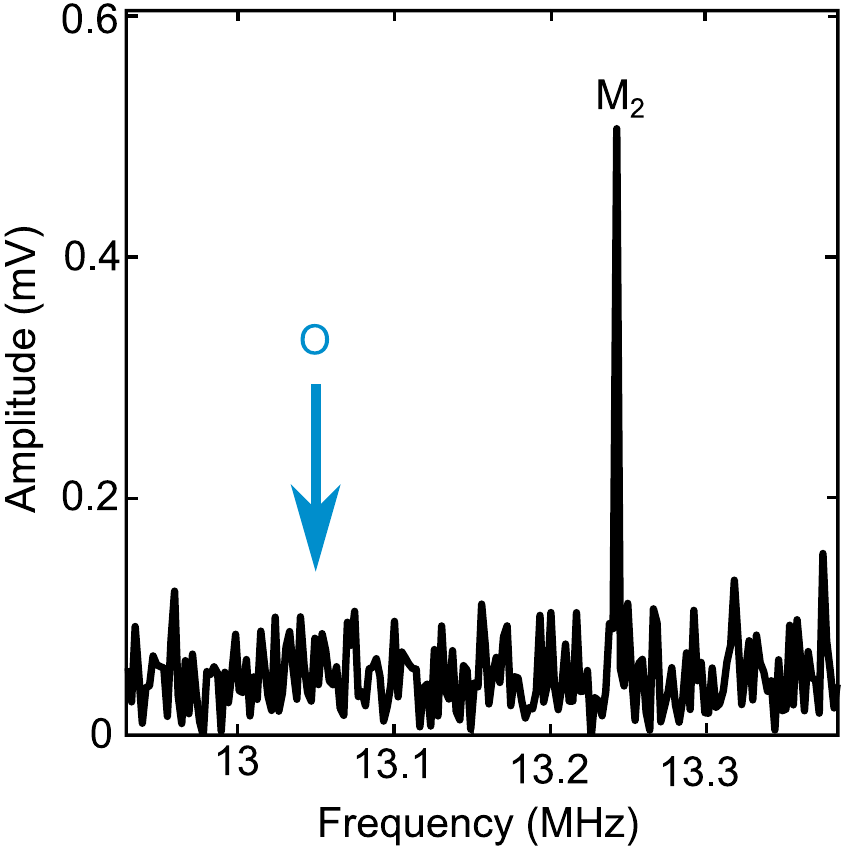}
	\caption{Response of the second spatial harmonic out-of-plane eigenmode M$_2$ appearing 200\,kHz higher in frequency than the  overtone O of the fundamental flexural out-of-plane eigenmode M.}
	\label{fig:af2}
\end{figure}

A relevant mechanism generating a signal at twice the driving force frequency for $\omega_d \approx \omega_0$ is the nonlinear coupling of the primary  mode M and the second spatial harmonic of the vibrations transverse to the nanostring, the eigenmode M$_2$. This mode appears at  frequency $\omega_2/2\pi=13.2$\,MHz, as shown in Fig.\,\ref{fig:af2}. It is almost $200$\,kHz above the overtone of the fundamental mode M, indicated by the blue arrow and labeled by O. This difference is a result of the non-negligible bending rigidity of the high-tension nanobeam under investigation, and thus the deviation from pure string-like behavior.
%						 FIGURE A2

The mode M$_2$ could be excited by the drive at frequency $\omega_d\approx\omega_0$. If the coordinate of this mode is $q_2$, the potential of the nonlinear coupling of this mode to the main-tone mode M has a term $U_{12}=\beta_{12}q^2q_2$, where $q$ is the coordinate of the  mode M. If the internal nonlinearity of the mode M$_2$ is disregarded, its equation of motion reads
\[\ddot q_2 + \omega_2^2 q_2 = -\beta_{12}q^2.\]
The forced vibrations of this mode induced by the forced vibrations $q(t)\approx q^{(1)}(t) = A\cos(\omega_d t+\varphi)$ of the mode M are described by the expression
\begin{align}
\label{eq:mode_2_vibrations}
q_2(t)\approx \frac{1}{2}\beta_{12}\frac{A^2}{4\omega_d^2-\omega_2^2}\cos(2{}\omega_d t+2\varphi)+ \mathrm{const}
\end{align}
The denominator in this expression is small compared to the denominator in the expression (\ref{eq:second_harmonic}) for the overtone of the main tone, $|4\omega_d^2-\omega_2^2|\ll \omega_0^2$. However, it is important that the parameter $\beta_{12}$ would be equal to zero in a symmetric nanoresonator. Moreover, it remains small in an asymmetric resonator, as the coupling results only from the distortion of the modes compared to the conventional sinusoidal shape. In addition, our measurement scheme effectively averages out the signal from the mode M$_2$.

%%%%%%%%%%%%%%%%%%%%%%%%%%%%%%%

\subsection{The effect of a nonlinear coupling to the driving force}

\subsubsection{Driving at $\omega_d \approx \omega_0$}

Another mechanism generating a signal at twice the drive frequency can be understood by recalling that the force on the nanostring under dielectric driving \cite{Unterreithmeier2009} comes from modulating the potential of the surrounding electrodes, see Fig.~\ref{fig:f1}. The nanostring is a part of the capacitor formed by these electrodes. The force on the mode with a coordinate $q$  is $\frac{1}{2}(\partial C/\partial q) V^2$, where $C$ is the capacitance and $V$ is the potential applied to the electrodes. This potential has an RF part that oscillates at the drive frequency $f_d$, $V_\mathrm{RF} = V_d\cos\omega_dt$, and a (usually large) DC part $V_\mathrm{DC}$. Therefore the force is periodic with period $2\pi/\omega_d$, but since the force as a whole is $\propto V^2$, it  has terms that oscillate not just at $f_d$, but also at $2{}f_d$. 

The force emerges only where  $\partial C/\partial q$ is nonzero, which in turn occurs where the system lacks inversion symmetry (on the contrary, if the electrodes formed a parallel-plate capacitor and the dielectric nanobeam was located symmetrically in the middle of the capacitor, we would have $\partial C/\partial q=0$). In the case of dielectric driving that we study, the system does not have inversion symmetry, and therefore the force does have a component at $2{}f_d$. This force is much smaller than the force at $f_d$ for $V_d \ll V_\mathrm{DC}$, but can become sizeable for large $V_d$.

We write the dielectric force component at twice the drive frequency as $\lambda F_d^2\cos 2{}\omega_d t$; the parameter $\lambda$ is determined by $\partial C/\partial q$. The vibrations caused by this force have the form
\begin{align}
\label{eq:2nd_overtone_direct}
q^{(2)}_\lambda (t) = \mathrm{Re}\,\frac{\lambda F_d^2\exp(2i\omega_d t)}{\omega_0^2-4\omega_d^2 + 4i\Gamma\omega_d}
\end{align}
Clearly, they contribute to the signal observed at $2{}f_d$. For close to resonance driving, $|\omega_0-\omega_d|\ll \omega_d$ the denominator in the above expression becomes $\approx -3\omega_0^2$.

\begin{figure} [t!]
	\centering
	\includegraphics[width=0.7\linewidth]{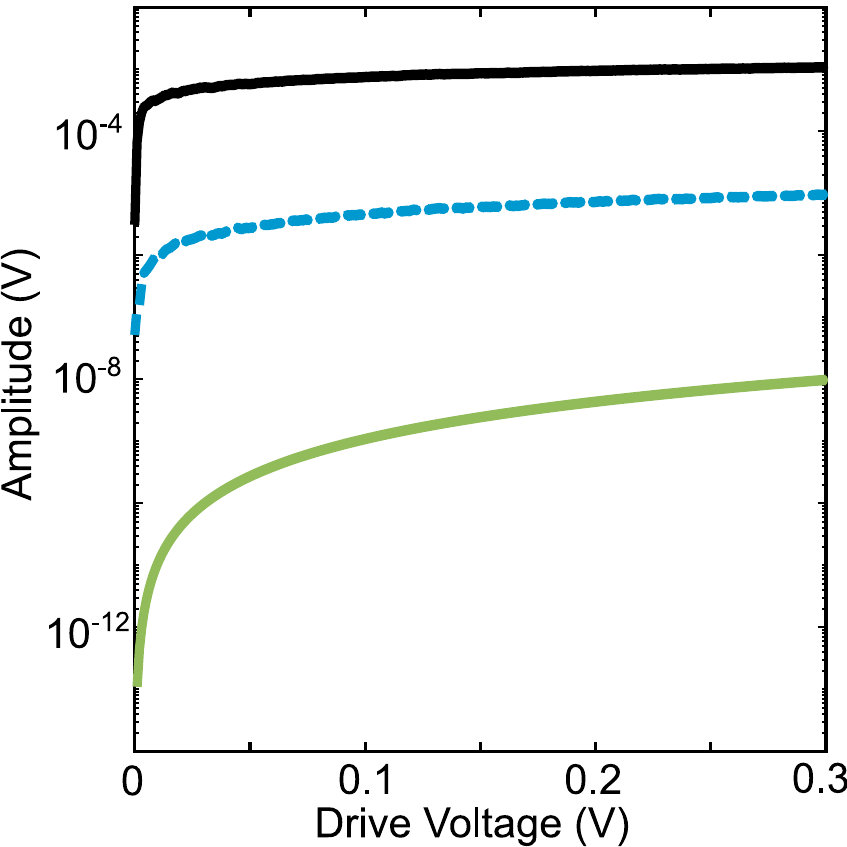}
	\caption{Influence of the nonlinear coupling to the driving force on the signal at $2{}f_d$. The green line shows the amplitude of vibrations at $2f_d$, which result from the nonlinear component of the drive and are described by Eq.~(\ref{eq:2nd_overtone_direct}). The black and dashed blue lines are taken from Fig.~\ref{fig:f4} and show, respectively, the measured amplitudes of the fundamental mode M and the  signal at  $2{}f_d$  (the amplitude of the signal at $2f_d$  has been rescaled to account for the different displacement-to-voltage conversion factor at this frequency). The results demonstrate a negligible influence of nonlinear driving.
	}
	\label{fig:af3}
\end{figure}

For our experiment, we find $\lambda = 5.88 \cdot 10^{-6}$\,s$^2/$m. Figure~\ref{fig:af3} plots the theoretical contribution of Eq.~(\ref{eq:2nd_overtone_direct}) to the signal at $2{}f_d$ in comparison to the experimentally determined amplitudes of the fundamental  mode M and the overtone signal O. For example, for a drive with $V_d=100$\,mV, we find a contribution which is more than two orders of magnitude smaller than the measured signal at $2{}f_d$, suggesting only a minor role of the mechanism described by Eq.~(\ref{eq:2nd_overtone_direct}).

%						 FIGURE A3

The force from modulating the capacitance by the vibrations of the nanobeam also contains the parametric driving term $\lambda'qF_d\cos\omega_d t$. This term comes from the derivative $\partial^2 C/\partial q^2$ evaluated at the equilibrium position of the nanobeam. Because of this term, the mode vibrations $q^{(1)}(t) = A\cos(\omega_d t+\varphi)$ lead  to vibrations at twice the drive frequency, with the displacement of the form
\begin{align}
\label{eq:nonlinear_suscept}
q^{(2)}_{\lambda'} (t) =\frac{1}{2} \mathrm{Re}\,\frac{\lambda'A F_d\exp(2i\omega_d t+\varphi)}{\omega_0^2-4\omega_d^2 + 4i\Gamma\omega_d}.
\end{align}
Here, again, for resonant driving the denominator becomes $\approx -3\omega_0^2$.

\subsubsection{Resonant excitation, $\omega_d \approx \omega_0/2$}
\label{resonant}

The vibrations at $2{}\omega_d$ can be excited resonantly (see Fig.~\ref{fig:f6}). This is a mechanical analog of the resonant second harmonic generation in nonlinear optics. It occurs if the mode lacks inversion symmetry and is driven close to a half of its eigenfrequency, i.e., $|2{}\omega_d -\omega_0|\ll \omega_0$. The cubic nonlinearity of the potential of the mode M contributes to this resonant excitation, as described in Sec.~\ref{results}. The mechanisms of nonlinear driving discussed in this Appendix contribute to the effect as well. We now consider these latter contributions. They are additive and therefore can be analyzed separately.

The effect of the  direct nonlinear drive is described by Eq.~(\ref{eq:2nd_overtone_direct}) with $\omega_d$ close to $\omega_0/2$. 
Therefore the denominator in Eq.~(\ref{eq:2nd_overtone_direct}) is small, a signature of the resonant second-harmonic generation.
In addition, forced vibrations at the angular frequency $\omega_d$ also resonantly excite vibrations at $2{}\omega_d\approx \omega_0$ via the nonlinear (parametric) coupling to the force. 
This contribution is described by Eq.~(\ref{eq:nonlinear_suscept}) in which one should set $A=F_d/(\omega_0^2-\omega_d^2)$ and $\varphi=0$. 
Again, the denominator in Eq.~(\ref{eq:nonlinear_suscept}) becomes resonantly large for $\omega_d$ close to $\omega_0/2$. However, if the effect of nonlinear driving is small for the driving at frequency $\approx \omega_0$, it is expected to be small for the driving at frequency $\omega_0/2$.

Importantly, the mode M$_2$ is no longer close to resonance with the overtone of the drive frequency. Given the weakness of the coupling to this mode, excitation of its vibrations can be disregarded. We expect therefore that, in our system, the resonant second harmonic generation is due to the internal cubic nonlinearity of the mode potential $U(q)$.

%%%%%%%%%%%%%%%%%%%%%%%%%%%%%%%%%%%%%%%%%%%%%%%%%%%%%%%%%%%%%%%%%%
%%%%%%%%%%%%%%%%%%%%%%%%%%%%%%%%%%%%%%%%%%%%%%%%%%%%%%%%%%%%%%%%%%
%%%%%%%%%%%%%%%%%%%%%%%%%%%%%%%%%%%%%%%%%%%%%%%%%%%%%%%%%%%%%%%%%%

\section{DC voltage dependence of nonlinearity parameters $\beta$ and $\gamma$}
\begin{figure} [t!]
	\centering
	\includegraphics[width=0.7\linewidth]{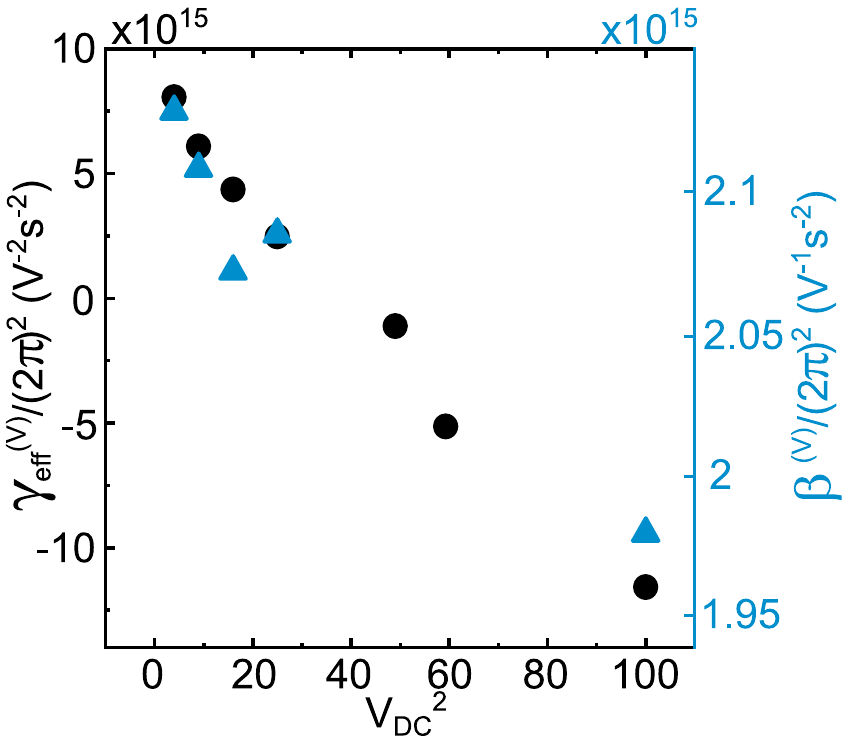}
	\caption{DC voltage dependence of the nonlinearity parameters. Both $\beta^{(V)}$ and $\gamma_\mathrm{eff}^{(V)}$ are plotted against the square of the applied DC voltage. The sign of $\gamma_\mathrm{eff}^{(V)}$ changes from positive to negative near $V_\mathrm{DC}=6$\,V. The cubic nonlinearity $\beta^{(V)}$ has been determined using the procedure described in Sec.~\ref{results}. It is only accessible in the region of not too small $|\gamma_\mathrm{eff}^{(V)}|$.
	}
	\label{fig:af5}
\end{figure}

To explore in more detail how the broken symmetry arises for the dielectrically controlled nanostring, we performed a series of measurements at different DC voltages $V_\mathrm{DC}$ between $0$ and $10$\,V. These measurements enable finding the DC voltage dependence of the two nonlinearity parameters $\beta^{(V)}$ and $\gamma_\mathrm{eff}^{(V)}$. The result of the experiment is shown in Fig.\,\ref{fig:af5}. A quadratic dependence of the both parameters on $V_\mathrm{DC}$ is observed, which confirms the dielectric nature of the nonlinearities. The effective Duffing parameter $\gamma_\mathrm{eff}^{(V)}$ changes sign near a DC voltage of $6$\,V, indicating the transition from stiffening to softening behavior.
The value of $\beta^{(V)}$ is only accessible in the region where  $\gamma_\mathrm{eff}^{(V)}$ is not very small, as close to $\gamma_\mathrm{eff}^{(V)}=0$ we observe the zero-dispersion regime \cite{Dykman1990d, Soskin2003,Huang2019} where we cannot extract $\beta^{(V)}$ using the procedure described in the main text. 

It is seen from Fig.~\ref{fig:af5} that the change of $\beta$ in the considered range of the DC voltage is only $\sim 10\%$ whereas $\gamma_\mathrm{eff}$ changes very significantly. The small change of $\beta$ is qualitatively consistent with the relative change of the eigenfrequency by $\lesssim 5\times 10^{-3}$ that we have observed in the same range. The strong relative change of $\gamma_\mathrm{eff}$ can be a result of the small value of this parameter as it goes through zero with the varying $V_\mathrm{DC}$ in the considered range.

%						 FIGURE A5
%
%
%%%%%%%%%%%%%%%%%%%%%%%%%%%%%%%%%%%%%%%%%%%%%%%%%%%%%%%%%%%%%%%%%%	 	

\section{Theory of forced vibrations in terms of the action-angle variables}
\label{app:theory}

Here we use the action and angle variables $I$ and $\phi$ to calculate the amplitude of the stable vibrational state of a resonantly driven mode with cubic and quartic nonlinearity with the potential energy $U(q)$.
Since the decay rate of the mode is small, $I$ and $\phi-\omega_d t$ remain almost constant over the drive period $2\pi/\omega_d$. Therefore the right-hand sides of Eq.~(\ref{eq:eom_I_phi_full})  for $\dot I$ and $\dot\phi$ can be averaged over the vibration period. Formally, we can define the averaging as 
\[\overline{L(I,\phi)} = (2\pi)^{-1}\int_0^{2\pi}d\phi \,L(I,\phi).\]
Averaging of the terms that contain the time-dependent factor $\cos(\omega_d t)$ can be done by writing this factor as $\cos[(\omega_d t-\phi) +\phi)$ and averaging over $\phi$ for a given $\omega_dt-\phi$. In the stationary state $\overline{\phi} = \omega_d t+\varphi_\mathrm{st}$.

Taking into account the explicit form of the function $R$ in Eq.~(\ref{eq:working}), we write this equation for the stationary values $\overline{I}=I_\mathrm{st},  \overline{\phi}-\omega_d t= \varphi_\mathrm{st}$ as 
\begin{align}
\label{eq:stationary_explicit}
	 - 2 \Gamma I\st - \frac{1}{2} {}A_1(I\st) F_d \sin\varphi\st &=0, \nonumber\\
	 \omega(I\st) - \frac{1}{2}  {\left.  \frac{ \partial {}A_1(I)}{\partial I} \right|}_{I\st} F_d \cos\varphi\st &=\omega_d    \, ,
\end{align}
where ${}A_1(I)$ is the amplitude of the term in the periodic function $q(I,\phi)$ that oscillates like $\cos\phi$, i.e., the term $q^{(1)} = {}A_1(I)\cos\phi$ in the expansion $q=\sum_n {}A_n(I)\cos n\phi$.
It follows from Eq.~(\ref{eq:stationary_explicit}) that
\begin{align}
	\label{eq:action_explicit}
	{\left( \frac{4 \Gamma I\st} {{}A_1(I\st) F_d} \right)}^2
	+ 4{\left[ \frac{\omega(I\st)-\omega_d}{ {\left.  \frac{ \partial {}A_1(I)}{\partial I} \right|}_{I\st}  F_d }   \right]}^2 =1 \, .
\end{align}
We have used that $\overline{p \, \partial_I q}=0$, as well as $\overline{p \, q}=0$, since the function $q(I,\phi)$ is even in $\phi$ whereas the function $p(I,\phi)$ is odd. 

The action $I$ is a function of energy,  $I\equiv I(E)$. This function is monotonic and invertible. Therefore we can rewrite Eq.~(\ref{eq:action_explicit}) as the equation for the energy of the stable state. We express the stationary equation as a function of the energy
\begin{align}
	\label{eq:main}
	{\left( \frac{4 \Gamma I_E(E\st)} {A_{1E}(E\st) F_d} \right)}^2
	+
	{\left( \frac{\omega_E(E\st) -\omega_d}{\frac{1}{2}  \omega_E(E\st) {\left. \frac{ \partial A_{1E}(E)}{\partial E} \right|}_{E\st} F_d  }  \right)}^2
	=
	1 \,,
\end{align}
Here we use the subscript $E$ to indicate that the corresponding parameter is considered as a function of energy $E$, not action $I$.

\begin{figure}[t!]
	\centering
	\includegraphics[width=0.7\linewidth]{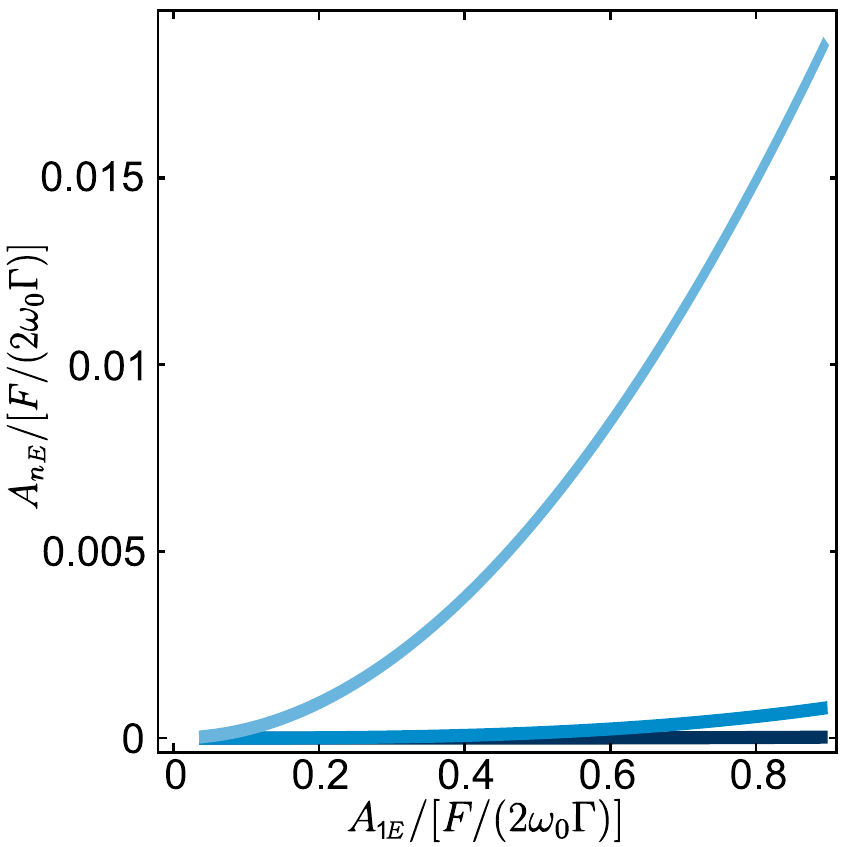}
	\caption{Amplitudes $A_{nE}$ of the overtones at frequencies $n\omega_E$ with $n=2,3,4$. The amplitudes are calculated from Eq.~(\ref{eq:overtones}) by varying  the energy $E$ and, respectively, the amplitude $A_{1E}$ of the main tone. For convenience of the comparison with the experiment, the amplitudes are scaled by the drive-dependent factor, but in fact they are calculated from Eqs.~(\ref{eq:periodic_displacement}) and (\ref{eq:overtones}) in the absence of driving. Shown are the amplitudes 
	$A_{2E}= 2|a_2|$ (blue),  
 $A_{3E}= 2|a_3|$ (mid blue), and $A_{4E}= 2|a_4|$ (dark blue)
   }
	\label{fig:af4}
\end{figure}

The turning points $q_L$ and $q_R$ for the motion in the single-well potential $U(q)$ with energy $E$ are given by the 
two real roots of the quartic equation
\begin{align}
	\label{eq:energy}
	E = \frac{1}{2} \omega_0^2 q_{\nu}^2+  \frac{1}{3} \beta q_{\nu}^3+  \frac{1}{4} \gamma q_{\nu}^4 \, , \quad  \, \nu=L, R \, ,
\end{align}
with $q_R>0$ and $q_L<0$. 
The other two complex conjugate roots are denoted as $q_3 \pm i q_4$.
Then we introduce the two coefficients $B_{\nu}$ with $\nu=L,R$  given by
\begin{align}
	B_{\nu} &= \left| q_{\nu}- q_3 - i q_4 \right| \, .
\end{align}
and the parameter
\begin{align}
	k &=  \sqrt{ 
			\frac{ {\left( q_R - q_L \right)}^2 - {\left( B_R - B_L\right)}^2  }{4 B_R B_L} }\, .
\end{align}
Then  the frequency $\omega_E(E)$ is given by
\begin{align}
	\omega_E(E) = \frac{\pi}{2} \sqrt{\frac{\gamma}{2}} \frac{ \sqrt{B_L B_R}  }{ 
	K	}  \,,
\end{align}
with the 
complete elliptic integral of the first kind $K =  \mbox{K} \left[  k \right]$.
The exact solution of the motion at given energy $E$ reads
\begin{align}
	\label{eq:periodic_displacement}
	 q_E (\tau) &= \frac{B_R q_L- B_L q_R }{ B_R  - B_L   }
	  \nonumber \\
	&+ \frac{ 2 B_R B_L }{  {\left( B_R  - B_L \right)}^2 }
	 \frac{q_R -  q_L }{ \frac{B_R  +B_L}{ B_R  - B_L }
	 +  \mbox{cn}_k\left( \tau \right)}
	\, .
\end{align}
with the scaled time $\tau =\sqrt{( \gamma/2 ) B_LB_R} \, t $ and the Jacobi elliptic cosine function $ \mbox{cn}\left( \tau \right)$.
The Fourier components of the motions are 
\begin{align}
	\label{eq:overtones}
	a_n = \frac{1}{4K} \int^{4K}_0 \!\!\!\! d\tau \,\, q_E(\tau) \,\, e^{-i n \frac{\pi}{2K} \tau}
\, .		
\end{align}
An example is shown in Fig.~\ref{fig:af4} in which we vary the energy $E$ in a given range and plot parametrically the components $A_{2E}(E)= 2|a_2|$, $A_{3E}(E)= 2|a_3|$ and $A_{4E}(E)= 2|a_4|$ 
as functions of the component $A_{1E}(E) =  2|a_1|$. We recall that, in the stationary state $A_{1E}$ gives the amplitude of the signal at the drive frequency $\omega_d$, whereas $A_{2E}$ gives the amplitude of the signal at frequency $2\omega_d$.

The action as a function of the energy is  
\begin{align}
	I_E(E) &=
	\frac{1}{2\pi}
\oint  \!  dq \,\, \sqrt{ 2\left[E - U(q) \right] } = 
\frac{\omega(E)}{2\pi} \int^{2\pi}_0 \!\!\!\!\! d\varphi \, {\left( \frac{d q}{d \varphi} \right)}^2 \, , \\
&= 2 \, \omega_E(E) \, \sum_{n=1}^{\infty} n^2 {\left| a_n\right|}^2
	 \, .
\end{align}
The explicit expression of the Fourier components $a_n$ were
calculated in \cite{Dykman1991a} for the potential $U_B(q)$ of  Eq.~(\ref{eq:bias}).
 
For small energies counted off from the minimum of the potential one can approximate $I_E(E) \approx E/\omega_0$ and $A_{1E}(E) \approx \sqrt{2 E/\omega_0^2}$ and 
expand the frequency $\omega_E$ to the second order in $E$. Such an approximation works in the case $|\gamma_\mathrm{eff}| \ll \gamma$, where the linear in $E$ term in $\omega_E$ is comparatively small. One then obtains from Eq.~(\ref{eq:main}) an explicit equation for the vibration amplitude in the stationary state $A\equiv A_{1E}(E_\mathrm{st})$ 
\begin{align}
	\label{eq:anal}
	A^2
	\left[ \Gamma^2
	+
	{\left( \omega_0 + \frac{3   \gamma_\mathrm{eff} }{8\omega_0} A^2 +   \frac{ \chi \gamma^2}{\omega_0^3} A^4
	-\omega_d \right)}^2
	\right]  \approx \frac{F_d^2}{4 \omega_0^2}
	 \, ,
\end{align}
with the parameter
\begin{align}
	\chi =\frac{69}{256} 
	{\left[
	\frac{102}{23} {\left( 1 - \frac{\gamma_\mathrm{eff} }{\gamma} \right)} - \frac{171}{115}  {\left( 1 - \frac{\gamma_\mathrm{eff} }{\gamma} \right)}^2 -1 
	\right]}
	 \, .
\end{align}
Within the range of the applied force and detuning, and for the parameters of the system studied in the experiments 
the solutions of Eq.~(\ref{eq:anal}) accurately reproduce the solution of the full equation  (\ref{eq:main}).

We note that the standard RWA breaks down as soon as the term $\propto A^4$ in the brackets in Eq.~(\ref{eq:anal}) becomes relevant.
This occurs when  $A^2 \sim  |\gamma_\mathrm{eff}| \omega_0^2 / \gamma^2$ as stated in the main text.
In the range of parameters studied in the experiment, Eq.~(\ref{eq:anal}) can have one or three real solutions, as is the case also for the standard Duffing equation. However, the dependence of these solutions on the frequency of the drive is significantly different from that for the Duffing model.
%
%

%\bibliography{md_bib_modified}
		
	%

\end{document}